\documentclass[aps,twocolumn,showpacs,preprintnumbers,amsmath,amssymb,superscriptaddress,prc]{revtex4-2}

\usepackage{graphicx}
\usepackage{dcolumn}
\usepackage{bm}

\begin{document}


\title{Determination of the spins and parities for the 0$_{4}^{+}$ and 0$_{5}^{+}$ states in $^{100}$Zr}

\author{J.~Wu}

\affiliation{%
National Nuclear Data Center, Brookhaven National Laboratory, Upton, NY 11973, USA
}%
\affiliation{%
Physics Division, Argonne National Laboratory, Lemont, IL 60439, USA
}%

\author{M.P.~Carpenter}
\affiliation{%
Physics Division, Argonne National Laboratory, Lemont, IL 60439, USA
}%

\author{F.G.~Kondev}
\affiliation{%
Physics Division, Argonne National Laboratory, Lemont, IL 60439, USA
}%

\author{R.V.F.~Janssens}
\affiliation{%
Department of Physics and Astronomy, University of North Carolina at Chapel Hill, Chapel Hill, North Carolina 27599, USA
}%
\affiliation{%
Triangle Universities Nuclear Laboratory, Duke University, Durham, North Carolina 27708, USA
}%

\author{S.~Zhu}
\thanks{Deceased.}
\affiliation{%
Physics Division, Argonne National Laboratory, Lemont, IL 60439, USA
}%

\author{E.A.~McCutchan}
\affiliation{%
National Nuclear Data Center, Brookhaven National Laboratory, Upton, NY 11973, USA
}%

\author{A.D.~Ayangeakaa}
\affiliation{%
Department of Physics and Astronomy, University of North Carolina at Chapel Hill, Chapel Hill, North Carolina 27599, USA
}%
\affiliation{%
Triangle Universities Nuclear Laboratory, Duke University, Durham, North Carolina 27708, USA
}%

\author{J.~Chen}
\thanks{Current address: Facility for Rare Isotope Beams, Michigan State University, East Lansing, Michigan 48824, USA.}
\affiliation{%
Physics Division, Argonne National Laboratory, Lemont, IL 60439, USA
}%

\author{J.~Clark}
\affiliation{%
Physics Division, Argonne National Laboratory, Lemont, IL 60439, USA
}%

\author{D.~J.~Hartley} 
\affiliation{%
Department of Physics, United States Naval Academy, Annapolis, Maryland 21402, USA
}%

\author{T.~Lauritsen}
\affiliation{%
Physics Division, Argonne National Laboratory, Lemont, IL 60439, USA
}%

\author{N.~Pietralla}
\affiliation{%
Institut f\"ur Kernphysik, Technische Universit\"at Darmstadt, 64289 Darmstadt, Germany
}%

\author{G.~Savard}
\affiliation{%
Physics Division, Argonne National Laboratory, Lemont, IL 60439, USA
}%

\author{D.~Seweryniak}
\affiliation{%
Physics Division, Argonne National Laboratory, Lemont, IL 60439, USA
}%

\author{V.~Werner}
\affiliation{%
Institut f\"ur Kernphysik, Technische Universit\"at Darmstadt, 64289 Darmstadt, Germany
}%

\date{\today}

\begin{abstract}

Two 0$^{+}$ states at 1294.5 and 1774.0 keV, together with three 2$^{+}$ and one 4$^{+}$ levels, 
were identified or unambiguously spin-parity assigned for the first time in $^{100}$Zr utilizing  $\gamma$-ray spectroscopy and $\gamma$-$\gamma$ angular correlation techniques with the Gammasphere spectrometer, following the $\beta^{-}$ decay of neutron-rich, mass separated $^{100,100m}$Y isotopes. 
Comparisons with recent Monte Carlo Shell-Model (MCSM) calculations indicate that these two states are candidates for the bandhead of a sequence in a shape-coexisting spherical minimum predicted to be located around $\approx$1500 keV. According to the measured relative B(E2)$_{relative}$ transition probabilities, the 0$_{5}^{+}$ state exhibits decay properties which more closely align with those predicted for a spherical shape, while the 0$_{4}^{+}$ level is suggested to be associated with a weakly-deformed shape similar to one related to the 0$_{2}^{+}$ state. 
\end{abstract}

\maketitle


\section{Introduction}

Over the last several decades, experimental advances in studying neutron-rich nuclei have revealed that the shell structure established near the valley of stability changes as a function of proton-to-neutron ratio. These modifications in shell structure have been associated, in part, with the tensor contribution to the nucleon-nucleon interaction. Some of the experimental observations resulting from this interaction are the disappearance of some well-established magic numbers, and the occurrence of new ones in the neutron-rich regime~\cite{Otsuka20}. One recent example of this phenomenon occurs in the Ca isotopes, where the observation of an enhanced excitation energy for the first 2$^{+}$ state in $^{54}$Ca ($N$ = 34) is associated with a large shell gap at $N$ = 34 which develops between the $f_{5/2}$ and $p_{3/2}$ single-particle neutron orbitals as a result of the fact that the proton $f_{7/2}$ shell is empty at $Z$ = 20 \cite{Steppenbeck13}. In contrast, by adding only two more protons, and thus partially occupying the f$_{7/2}$  proton orbital, the $N$ = 34 shell gap disappears, as has been shown experimentally for $^{56}$Ti \cite{Janssens02,Liddick04}. In recent publications by the Tokyo group, this example of altered shell structure has been labeled as "Type-I" shell evolution \cite{Otsuka16}.

Coupled to this shell evolution is an associated change in the nuclear shape. In transitional regions, one observes shape coexistence, where low-lying excited states appear associated with a shape different from that of the ground state. It has been recently shown that the presence of shape coexistence can also be enhanced by the tensor force, where deformation causes a rearrangement in orbital occupancies which can, in some instances, result in a lowering in energy of the deformed states.  This modification of the nuclear single-particle orbitals by the tensor force within the same nucleus is referred to as "Type-II" shell evolution \cite{Tsunoda14,Kremer16}. In most instances of shape coexistence, one observes a rather smooth change in the relative energies of the coexisting shapes as a function of $N$ and $Z$. The neutron-rich Zr nuclei are anomalous in this respect as a rapid shape change between spherical and deformed ground states occurs between $^{98}$Zr ($N$ = 58) and $^{100}$Zr ($N$ = 60). This phenomenon has been known for some time, and was first observed by Khan et al. in both the Zr isotopes \cite{Khan77}.

\begin{center}
\begin{figure*}
\resizebox{2.0\columnwidth}{!}{\rotatebox{90}{\includegraphics[clip=]{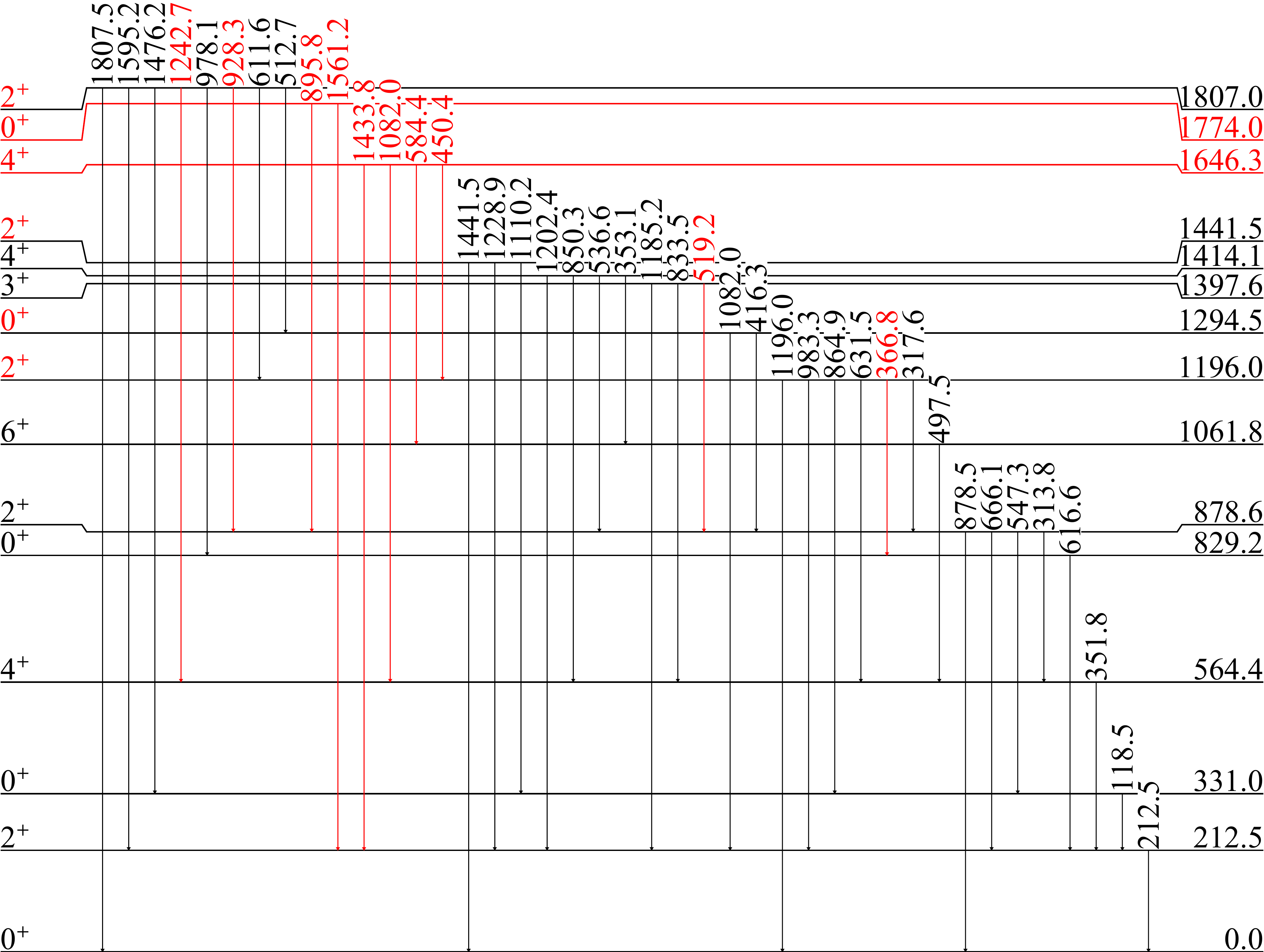}}}
\caption{\label{fig:ls} Partial level scheme of $^{100}$Zr observed in the $\beta$-decay of $^{100}$Y and $^{100m}$Y. The newly-identified levels, $\gamma$-ray transitions and assigned spins and parities are provided in red.} 
\end{figure*}
\end{center}

This apparent sudden change in shape has been interpreted as resulting from the strong proton-neutron tensor force between proton $\pi$g$_{9/2}$ and neutron $\nu$g$_{7/2}$ orbitals, where the enhanced occupation of the $\nu$g$_{7/2}$ state is due to an increased quadrupole deformation \cite{Otsuka16,Arseniev69,Sheline72}. Togashi et al \cite{Togashi16} have further suggested that this rapid change in the relative excitation energy of spherical and prolate structures, when going from $^{98}$Zr to $^{100}$Zr, fulfills the requirements of a first-order phase transition, based on the results of MCSM calculations which predict the excitation energy of the 0$_{4}^{+}$ spherical state to be $\approx$ 1.5 MeV in $^{100}$Zr \cite{Togashi16}. In contrast, IBM-CM calculations predict a configuration exchange between 0$^{+}_{1}$ and 0$^{+}_{2}$ states from $^{98}$Zr to $^{100}$Zr, with the spherical 0$^{+}_{2}$ level located at $\approx$ 300 keV in $^{100}$Zr \cite{Gavr22,Gavr20,Gavr19}. Thus, to test the assertion of a first order phase transition experimentally requires the determination of the excitation energy of the spherical 0$^{+}$ bandhead in $^{100}$Zr, and more generally, a more extensive knowledge of 0$^{+}$ excitations in this nucleus. Here, we report on the observation of two new excited 0$^{+}$ states in $^{100}$Zr, where spin and parity quantum numbers were unambiguously determined using the unique angular correlation pattern provided by 0$^{+}$ $\rightarrow$ 2$^{+}$ $\rightarrow$ 0$^{+}$ cascades, as measured with the Gammasphere spectrometer. In addition, the analysis has allowed for the identification of several new excited states relative to the previously known level scheme \cite{Singh21}, as well as for the determination of their spins and parities. 
The new data provide a stringent test of the MCSM calculations, as discussed below.

\begin{center}
\begin{figure}
\resizebox{1.0\columnwidth}{!}{\rotatebox{0}{\includegraphics[clip=]{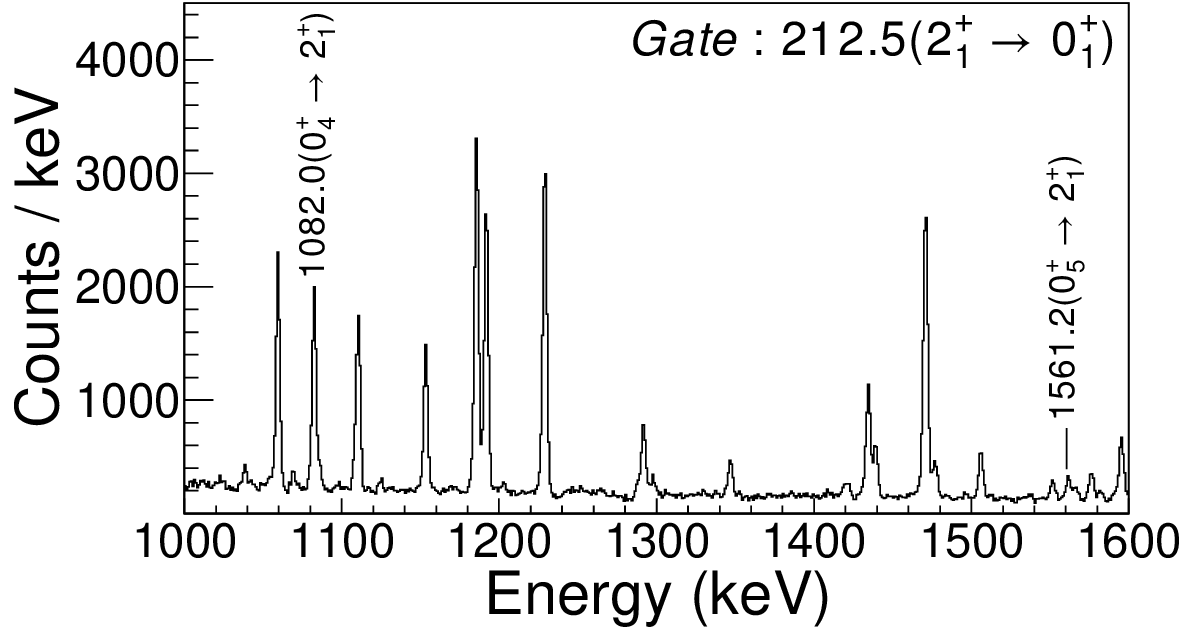}}}
\caption{\label{fig:spectrum} Coincidence spectrum for $^{100}$Zr following the $\beta$-decay of $^{100,100m}$Y gated on the 212.5-keV (2$^{+}_{1}$$\to$$0^{+}_{1}$) transition. The 1082.0-keV (0$^{+}_{4}$$\to$$2^{+}_{1}$) and 1561.2-keV (0$^{+}_{5}$$\to$$2^{+}_{1}$) transitions are identified.}  
\end{figure}
\end{center}

\begin{center}
\begin{figure}
\resizebox{1.0\columnwidth}{!}{\rotatebox{0}{\includegraphics[clip=]{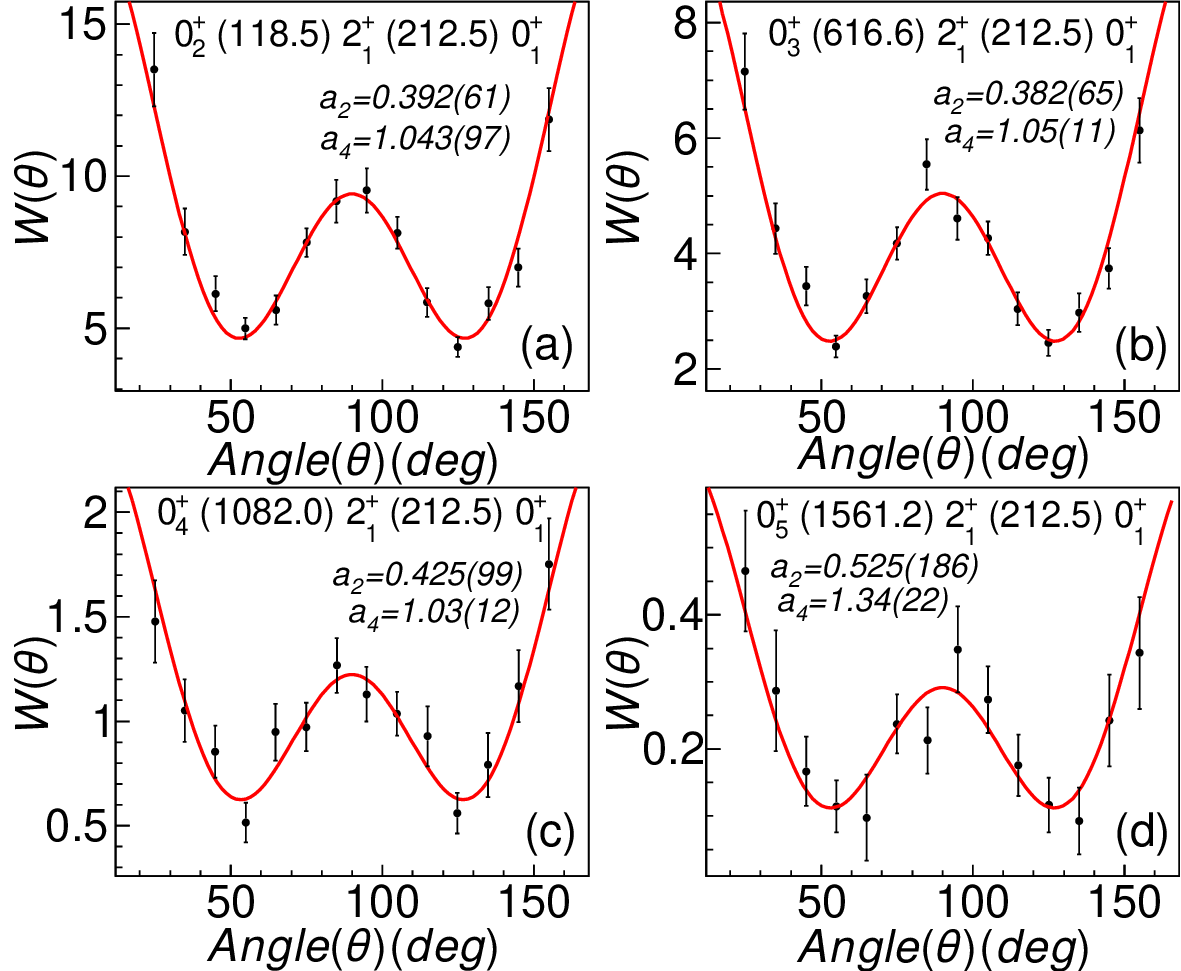}}}
\caption{\label{fig:corr1} The $\gamma$-$\gamma$ angular correlations for the (a) 118.5$-$212.5 keV, (b) 616.6$-$212.5 keV, (c) 1082.0$-$212.5 keV and (d) 1561.2$-$212.5 keV pairs of cascade $\gamma$-ray transitions.} 
\end{figure}
\end{center}

\makeatletter
\def\thickhline{%
  \noalign{\ifnum0=`}\fi\hrule \@height \thickarrayrulewidth \futurelet
   \reserved@a\@xthickhline}
\def\@xthickhline{\ifx\reserved@a\thickhline
               \vskip\doublerulesep
               \vskip-\thickarrayrulewidth
             \fi
      \ifnum0=`{\fi}}
\makeatother

\begin{table*}[ht!]
\flushleft

\caption{\label{tab:data} Relative Intensity ($I_{\gamma}$) and Branching Ratios ($BR_{\gamma}$) for the decay of selected excited states in $^{100}$Zr. The B(E2) ratios (B(E2)$_{relative}$) between 0$^{+}$ and 2$^{+}$ levels are reported with respect to the most intense transition decaying from each level. They are calculated based on the Branching Ratios by taking into account the Internal Conversion Coefficients calculated with BrIcc \cite{BrIcc}. The values of B(E2)$_{relative}$ tagged by asterisks (*) were calculated by assuming pure B(E2) transitions. The E0 branching from the 0$_{2}^{+}$ to 0$_{1}^{+}$ is obtained from the $\gamma$-$\gamma$ coincidence data.} 
\setlength{\tabcolsep}{8pt}
\renewcommand{\arraystretch}{1.2}
\begin{tabular*}{1.0\textwidth}{ c   c   c   c   c   c   c   c   c}
\hline\hline
$E_{initial} (keV) $ & $J_{i}^{\pi}$ & $E_{\gamma}$ (keV) & $E_{final}$ (keV) & $J_{f}^{\pi}$ & $I_{\gamma}$ & $BR_{\gamma}$ & $\delta$ & B(E2)$_{relative}$\\
\hline
212.5         &  2$_{1}^{+}$                                    &    212.5(1)     &      0           &    0$_{1}^{+}$       &  100.0(23)    &               &                      &             \\
331.0         &  0$_{2}^{+}$                                    &    118.5(1)     &      212.5       &    2$_{1}^{+}$       &  12.29(37)    &     100(3)    &                      &            \\
              &                                                 &                 &      0           &    0$_{1}^{+}$       &  7.50(62)     &     62(5)     &                      &          \\
564.4         &  4$_{1}^{+}$                                    &    351.9(1)     &      212.5       &    2$_{1}^{+}$       &  28.16(85)    &               &                      &            \\
829.2         &  0$_{3}^{+}$                                    &    616.6(2)     &      212.5       &    2$_{1}^{+}$       &  5.49(13)     &               &                      &             \\
878.6         &   2$_{2}^{+}$                                   &    547.3(2)     &      331.0       &    0$_{2}^{+}$       &  4.43(35)     &     29(2)     &                      &    0.8(1)    \\
              &                                                 &    666.1(2)     &      212.5       &    2$_{1}^{+}$       &  15.33(37)    &     100(3)    & 8.0$^{+357}_{-42}$   &    1.0        \\
              &                                                 &    878.5(2)     &      0           &    0$_{1}^{+}$       &  7.97(22)     &     52(2)     &                      &    0.13(2)    \\
1061.8        &  6$_{1}^{+}$                                    &    497.5(3)     &      564.4       &    4$_{1}^{+}$       &  2.10(7)      &               &                      &            \\
1196.0        &  2$_{3}^{+}$                                    &    317.6(8)     &      878.6       &    2$_{2}^{+}$       &  0.485(43)    &     8.4(6)    &                      &    64(5)*  \\
              &                                                 &    366.8(9)     &      829.2       &    0$_{3}^{+}$       &  0.216(51)    &     3.7(9)    &                      &    14(3)    \\
              &                                                 &    865.1(2)     &      331.0       &    0$_{2}^{+}$       &  5.33(17)     &     92(4)     &                      &    4.6(2)   \\
              &                                                 &    983.3(3)     &      212.5       &    2$_{1}^{+}$       &  2.17(21)     &     37(4)     & -1.0$^{+8}_{-4}$     &    0.5$^{+2}_{-4}$  \\
              &                                                 &    1196.0(2)    &      0           &    0$_{1}^{+}$       &  5.79(16)     &     100(3)    &                      &    1.0  \\
1294.5        &  0$_{4}^{+}$                                    &    416.3(7)     &      878.6       &    2$_{2}^{+}$       &  0.142(24)    &     7(1)      &                      &    8(1)  \\
              &                                                 &    1082.0(2)    &      212.5       &    2$_{1}^{+}$       &  2.11(10)     &     100(5)    &                      &    1.0    \\
1397.6        &  3$_{1}^{+}$                                    &    519.3(7)     &      878.6       &    2$_{2}^{+}$       &  0.27(1)      &     6(1)      &                      &            \\
              &                                                 &    833.6(7)     &      564.4       &    4$_{1}^{+}$       &  0.86(3)      &     18(1)     &                      &            \\
              &                                                 &    1185.2(2)    &      212.5       &    2$_{1}^{+}$       &  4.87(15)     &    100(4)     &                      &            \\
1414.1        &  4$_{2}^{+}$                                    &    354.1(2)     &      1061.8      &    6$_{1}^{+}$       &  0.18(2)      &     7(1)      &                      &            \\
              &                                                 &    535.6(1)     &      878.6       &    2$_{2}^{+}$       &  1.57(6)      &     61(3)     &                      &            \\
              &                                                 &    850.3(3)     &      564.4       &    4$_{1}^{+}$       &  2.56(8)      &    100(4)     &                      &            \\
              &                                                 &    1202.4(2)    &      212.5       &    2$_{1}^{+}$       &  0.33(3)      &    13(2)      &                      &            \\
1441.5        &  2$_{4}^{+}$                                    &    1110.3(4)    &      331.0       &    0$_{2}^{+}$       &  2.80(17)     &     51(4)     &                      &    0.86(9) \\
              &                                                 &    1228.9(1)    &      212.5       &    2$_{1}^{+}$       &  5.49(22)     &     100(4)    &  $<$-16.1 \& $>$5.8  &    1.0 \\
              &                                                 &    1441.8(3)    &      0           &    0$_{1}^{+}$       &  1.17(17)     &     21(3)     &                      &    0.10(2)  \\
1646.3        &  4$^{+}$                                        &    450.4(1)     &      1196.0      &    2$^{+}$           &  0.281(47)    &     15(3)     &                      &             \\
              &                                                 &    584.4(2)     &      1061.8      &    6$^{+}$           &  0.067(22)    &     4(2)      &                      &             \\
              &                                                 &    1082.0(1)    &      564.4       &    4$^{+}$           &  0.744(49)    &     41(3)     &                      &             \\
              &                                                 &    1433.8(1)    &      212.5       &    2$^{+}$           &  1.82(20)     &     100(11)   &                      &             \\
1774.0        &  0$_{5}^{+}$                                    &    895.8(2)     &      878.6       &    2$_{2}^{+}$       &  0.060(30)    &     13(7)     &                      &    2(1)  \\
              &                                                 &    1561.2(1)    &      212.5       &    2$_{1}^{+}$       &  0.45(3)      &     100(7)    &                      &    1.0   \\
1807.0        &  2$_{5}^{+}$                                    &    512.7(1)     &      1294.5      &    0$_{4}^{+}$       &  0.292(36)    &     22(3)     &                      &    67$^{+16}_{-9}$  \\
              &                                                 &    611.6(6)     &      1196.0      &    2$_{3}^{+}$       &  0.437(65)    &     35(5)     &                      &    44$^{+11}_{-7}$*   \\
              &                                                 &    928.3(1)     &      878.6       &    2$_{2}^{+}$       &  0.48(10)     &     39(8)     &                      &    6(2)*  \\
              &                                                 &    978.1(9)     &      829.2       &    0$_{3}^{+}$       &  0.59(4)      &     47(4)     &                      &    6(1)   \\
              &                                                 &    1476.2(4)    &      331.0       &    0$_{2}^{+}$       &  0.79(2)      &     64(3)     &                      &    1.0(2)  \\
              &                                                 &    1595.2(5)    &      212.5       &    2$_{1}^{+}$       &  1.244(55)    &     100(5)    &  -4.5$^{+23}_{-311}$ &    1.0   \\
              &                                                 &    1807.5(3)    &      0           &    0$_{1}^{+}$       &  1.032(91)    &     83(8)     &                      &    0.47(9)  \\

\hline\hline 
\end{tabular*}

\flushleft
  \label{tabN82}
  \end{table*}

\section{Experimental details}

Excited states in $^{100}$Zr were populated in the $\beta^{-}$-decay of $^{100}$Y (J$^{\pi}$=4$^{+}$, T$_{1/2}$=940 ms) and $^{100m}$Y (J$^{\pi}$=1$^{+}$, T$_{1/2}$=727 ms)~\cite{Nubase20}, produced by the CAlifornium Rare Isotope Breeder Upgrade (CARIBU) facility \cite{Savard08} at Argonne National Laboratory. The $^{100,100m}$Y ions were charge bred in an ECR source, re-accelerated up to 384 MeV by the ATLAS accelerator, and delivered to a 50 mg/cm$^{2}$ lead catcher foil placed at the center of the Gammasphere spectrometer \cite{gammasphere}, which was comprised of 92 Compton-suppressed high-purity Ge detectors for this experiment. A total of 4.0 $\times$ 10$^{8}$ $\beta$-delayed $\gamma$ coincidence events were recorded by Gammasphere. In the off-line analysis, the data were sorted into $\gamma$-$\gamma$ coincidence histograms. The $^{100}$Zr level scheme was constructed, based on singles events as well as coincidence relationships. The spin-parity assignments were aided by a $\gamma$-$\gamma$ angular correlation analysis. 
For this purpose, several $\gamma$-$\gamma$ matrices were constructed by constraining the angles between pairs of cascading $\gamma$ rays within a 20 to 160 degrees range in steps of 10 degrees. The $a_{2}$ and $a_{4}$ angular-correlation coefficients were extracted by fitting the data with the usual function $W(\theta)=A_{0}[1+a_{2}P_{2}(cos(\theta))+a_{4}P_{4}(cos(\theta))$]~\cite{Rose67}. 
Due to differences in the detector efficiencies and the anisotropic distribution of Ge detectors across the array, it is necessary to normalize counts measured as a function of the angle between detector pairs.  This was done using the 547-118.5 keV (2-0-2) sequence, as the angular correlation for a gamma-ray cascade with an intermediate 0$^{+}$ level is isotropic.  The J=2 and 0 spins for the initial and intermediate levels, respectively, are well established \cite{Singh21}, and the J=2 value of the initial level is further confirmed in our analysis (see Fig. 4).


\section{Results and Discussion}

The $^{100}$Zr partial level scheme is presented in Figure~\ref{fig:ls}. A sample $\gamma$-ray coincidence spectrum, produced by gating on the 212.5-keV, 2$_{1}^{+}$ $\to$ 0$_{1}^{+}$  transition is found in Figure~\ref{fig:spectrum}, where the identified 1082.0- (0$^{+}_{4}$$\to$$2^{+}_{1}$) and 1561.2-keV (0$^{+}_{5}$$\to$$2^{+}_{1}$) transitions are indicated. The J$^{\pi}$=0$^{+}_{1}$  ground state and the J$^{\pi}$=0$^{+}_{2}$ one at 331.0 keV, as well as the J$^{\pi}$=0$^{+}_{3}$ level at 829.2 keV,  were known from previous studies~\cite{Wohn86,Singh21}, and they are confirmed in the present data. The spin assignments to the 331.0-keV and 829.2-keV levels were also confirmed by the angular-correlation analysis, as displayed in Figures~\ref{fig:corr1}(a) and~\ref{fig:corr1}(b). The deduced $a_{2}$, $a_{4}$ correlation coefficients are in good agreement with the theoretical values of (0.357, 1.14) expected for a 0-2-0 sequence. The possible 1-2-0, 2-2-0, 3-2-0, 4-2-0 assignments can be excluded, since such transitions are characterized by a $a_{4}$ $<$ 0.6 coefficient. 
The 1646.3-keV and its de-exciting $\gamma$ rays were identified for the first time in the present work. This level is assigned J$^{\pi}$=4$^{+}$ quantum numbers based on the observed 1433.8- and 584.4-keV $\gamma$ rays to the 2$^{+}_{1}$ and 6$^{+}_{1}$ levels of the ground-state band.

\begin{center}
\begin{figure*}
\resizebox{2.0\columnwidth}{!}{\rotatebox{0}{\includegraphics[clip=]{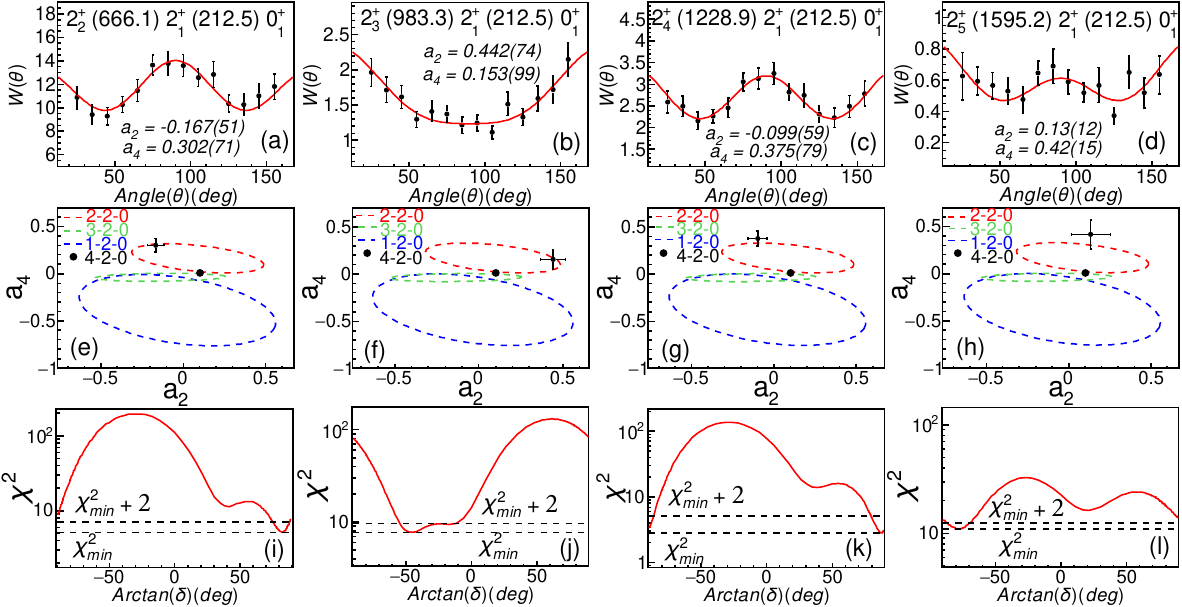}}}
\caption{\label{fig:corr2} (color online). (Top: a, b, c, d) Angular correlations for the 666.1$-$212.5, 983.3$-$212.5, 1228.9$-$212.5 and 1595.2$-$212.5 keV sequences fitted with the angular correlation function. (Middle: e, f, g, h) Determination of the spins of excited states through comparison with the parameters ($a_{2}$, $a_{4}$) for the possible spin sequences of 1-2-0 (blue), 2-2-0 (red), 3-2-0 (green), 4-2-0 (black dots). (Bottom: i, j, k, l) Calculated $\chi^{2}$ values obtained when varying the mixing ratios $\delta$ for the transition in 2-2-0 cascades, as a function of arctan($\delta$). The horizontal black dashed lines correspond to the $\chi^{2}_{min}$ + 2 and $\chi^{2}_{min}$ values.}  
\end{figure*}
\end{center}

The 1441.5- and 1807.0-keV levels were previously assigned J$^{\pi}$=(1,2$^{+}$)~\cite{Singh21}, but the angular correlation data in the present study clearly favor J$^{\pi}$=2$^{+}$. The fitted $\gamma$-$\gamma$ angular correlations for the 1228.9$-$212.5 keV and 1595.2$-$212.5 keV pairs, together with the 666.1$-$212.5 keV and 983.3$-$212.5 keV ones, are displayed in Fig.~\ref{fig:corr2}  (Top). The mixing ratios, $\delta(E2/M1)$, for the 2$^{+}$ $\to$ 2$^{+}$ transitions were determined through a $\chi^{2}_{min}$ analysis, as illustrated in  Fig.~\ref{fig:corr2} (Bottom), with  $2\sigma$ uncertainty (95\% confidence level) estimated as $\chi^{2}$=$\chi^{2}_{min}$+2~\cite{Tanabashi18}. 

The 1294.5-keV level was known previously with a tentative J$^{\pi}$=(2$^{-}$,3) assignment, based on the observed decay to 2$^{+}$ levels, the absence of $\gamma$ decay to 0$^{+}$ levels, and a possible feeding from the 1$^{-}$ parent \cite{Singh21}. The two gamma transitions decaying to the 2$^{+}_{1,2}$ states were confirmed by the present measurements and no transitions to 0$^{+}$ states were observed. The 1294.5-keV level was unambiguously assigned as J$^{\pi}$=0$_{4}^{+}$, based on the 1082.0$-$212.5 keV angular-correlation analysis, as seen in Figure~\ref{fig:corr1}(c).
The 1294.5-keV level was tentatively assigned J$^{\pi}$=(2$^{+}$) in the recent work of Kalaydjieva et al. \cite{Kalay23} by claiming a transition of 1295-keV to the ground state but lack of experimental evidence. The existence of this transition is doubtful as it is neither observed in this work nor reported in the previous work of Wohn et al. \cite{Wohn86}.
The 1774.0-keV state, and its de-exciting $\gamma$ rays, were identified for the first time in the present work. It is assigned J$^{\pi}$=0$^{+}_{5}$, based on the 1561.2$-$212.5 keV angular-correlation analysis, as displayed in Figure~\ref{fig:corr1}(d). It is worth noting that the newly-identified 0$_{4}^{+}$ (1294.5-keV) and 0$_{5}^{+}$ (1774.0-keV) levels were observed in an excitation energy range close (within 200-keV) to the MCSM 0$_{4}^{+}$ (spherical) level ($\approx$ 1500 keV) \cite{Togashi16,Gavr22,Gavr20,Gavr19}.

The systematic trends of the 0$^{+}_{1-5}$ states in $^{96-102}$Zr are displayed as a function of mass number in Fig.~5. The 0$_{2}^{+}$ state drops with increasing mass number from $^{96}$Zr reaching a minimum value of 331.0-keV at $^{100}$Zr then increasing again for $^{102}$Zr. Analogous to the 0$_{1,2}^{+}$ states in $^{98}$Sr \cite{Urban19}, the proximity of 0$_{2}^{+}$ and 0$_{1}^{+}$ levels in $^{100}$Zr indicates their weak interaction with little overlap in the wave functions, suggesting that the 0$_{2}^{+}$ state is associated with a collective structure different from that of the 0$_{1}^{+}$ ground state. This is consistent with the predicted prolate 0$^{+}_{1}$ and oblate 0$^{+}_{2}$ states in MCSM calculations \cite{Togashi16}. The "compressed" spectrum of five 0$^{+}$ levels below 2 MeV excitation energy in $^{100}$Zr is suggestive of a situation where, within a phase transition picture, states associated with different shapes coexist and mix to various degrees. As seen in Fig.~5, the 0$^{+}_{2,3,4}$ states decrease in excitation energy between $^{96}$Zr and $^{98}$Zr. A linear extrapolation to $^{100}$Zr then associates these three excited 0$^{+}$ levels in the $A$ = 96,98 isotopes with the ground state and the 0$^{+}_{2}$ and 0$^{+}_{3}$ states in the $A$ = 100 one, respectively. This striking correlation then also suggests that these levels are characterized by similar configurations, and are associated with similar shapes. Following this reasoning further tentatively associates the spherical 0$^{+}_{1}$ ground state of $^{96,98}$Zr with one of the two newly-identified 0$^{+}_{4,5}$ levels in $^{100}$Zr, implying an inversion between spherical and deformed states between $^{98}$Zr and $^{100}$Zr as predicted by the MCSM \cite{Togashi16} and IBM-CM calculations \cite{Gavr22,Gavr20,Gavr19}.


The properties of transitions for the levels of interest are given in Table I. The ratio of transition probabilities B(E2:2$_{2}^{+}$ $\to$ 0$_{2}^{+}$)/B(E2:2$_{2}^{+}$ $\to$ 0$_{1}^{+}$) has a fairly large value of 6.1(7), which indicates a sizable overlap of the wavefunctions of the 2$^{+}_{2}$ and 0$^{+}_{2}$ states. The fact that the energy of the transition linking the two levels is large (547.3 keV), when compared to the 212.5-keV 2$_{1}^{+}$ $\to$ 0$_{1}^{+}$ one, then suggests that the 0$^{+}_{2}$ state might be located in a potential minimum with a different shape than the ground-state one. Based on available calculations, the 0$^{+}_{2}$ level could possibly be associated with sphericity within the IBM-CM framework, or with the oblate deformation calculated within the MCSM. Based on the excitation energies of 2$^{+}$ states in neighboring spherical nuclei, a transition of the order of 1 MeV or more would be anticipated for the first excited level based on a spherical 0$^{+}_{2}$ band head. The actual 547.3-keV,  2$_{2}^{+}$ $\to$ 0$_{2}^{+}$ energy then argues for an excitation in a weakly-deformed minimum. An oblate structure for the 0$_{2}^{+}$ level has also been proposed by Garrett et al. \cite{Garrett22}. The previous work by Urban et al \cite{Urban19}, which had confirmed the placement of the 2$^{+}_{2}$ level on top of the 0$^{+}_{2}$ state, proposed an interpretation of the latter as a 2p-2h excitation of two neutrons in the 11/2$^{-}$[505] orbitals of h$_{11/2}$ parentage resulting in an oblate shape.

Similar arguments can be made for the 0$_{3}^{+}$ state, where the observed 366.8-keV, 2$_{3}^{+}$ $\to$ 0$_{3}^{+}$ transition has a much larger strength than those for decays toward the 0$_{1}^{+}$ and 0$_{2}^{+}$ levels. This observation would imply an excitation in another deformed minimum.

The present work establishes the 1807.0-keV, 2$_{5}^{+}$ level as the lowest 2$^{+}$ excitation feeding the 0$_{4}^{+}$ state with a 512.7-keV transition (Fig. 1). As can be seen from Table I, this transition dominates the deexcitation pattern out of the 2$_{5}^{+}$ level; i.e. the pattern mirrors that discussed above for the 2$_{2}^{+}$ $\to$ 0$_{2}^{+}$ case. In addition, the ratio B(E2:0$_{4}^{+}$ $\to$ 2$_{2}^{+}$)/B(E2:0$_{4}^{+}$ $\to$ 2$_{1}^{+}$) of 8(1) also argues for a similarity in the collective character of the structures based on the 0$_{4}^{+}$ and 0$_{2}^{+}$ states, and suggests the presence of weakly-deformed shapes in both cases.

Finally, the B(E2:0$_{5}^{+}$ $\to$ 2$_{2}^{+}$)/B(E2:0$_{5}^{+}$ $\to$ 2$_{1}^{+}$) ratio of 2(1) indicates comparable decay strengths toward the 2$_{1}^{+}$ and 2$_{2}^{+}$ levels. The lack of a preferred decay path toward states in potentials with different deformation can then be viewed as an indication for a different shape associated with the 0$_{5}^{+}$ state. Hence, based on this circumstantial evidence, this level is most likely the one corresponding to the predicted spherical shape in $^{100}$Zr.

\begin{center}
\begin{figure}
\resizebox{1.0\columnwidth}{!}{\rotatebox{0}{\includegraphics[clip=]{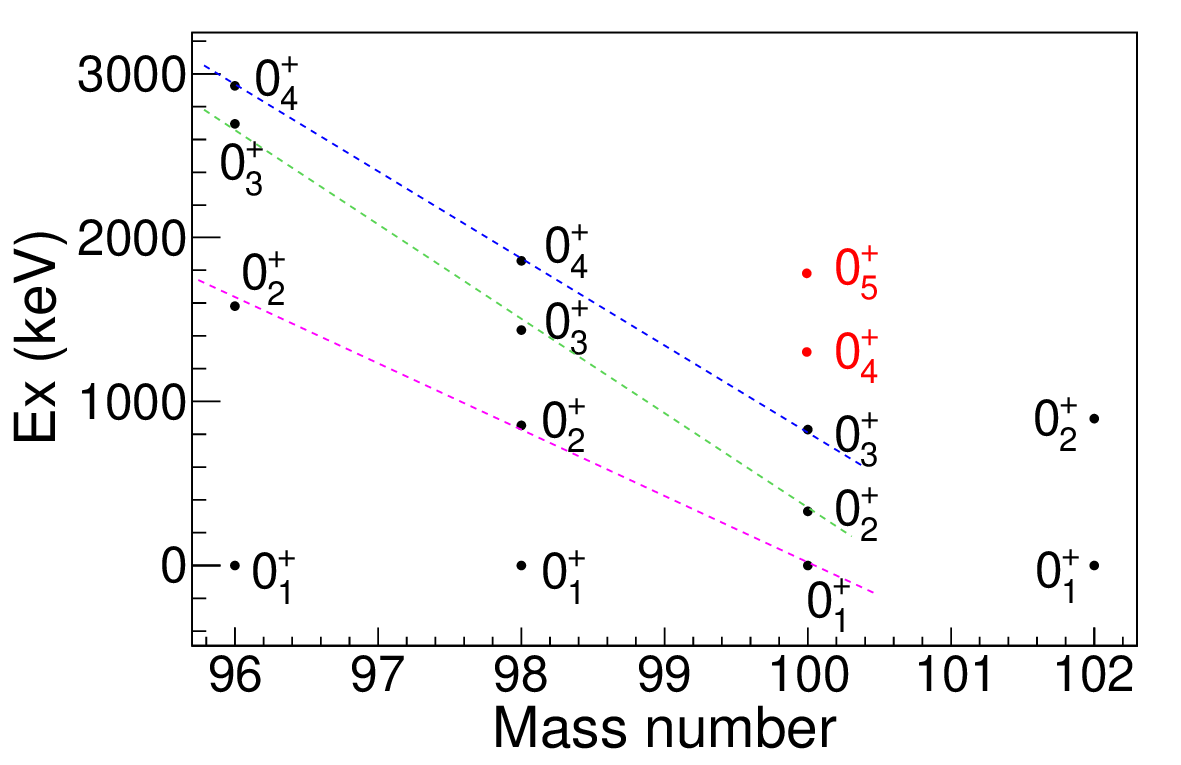}}}
\caption{\label{fig:chamber} Systematic trends of the level energies for the 0$^{+}$ states in $^{96-102}$Zr \cite{Singh21,Ariola08,Chen20}. The data points shown in red present the newly-identified 0$^{+}_{4,5}$ states in $^{100}$Zr.}
\end{figure}
\end{center}


\section{Summary}

The 0$^{+}_{4}$ and 0$^{+}_{5}$ states of $^{100}$Zr are identified for the first time by utilizing the unique 0-2-0 angular correlation signature, providing two candidates for the spherical 0$^{+}$ state predicted by MCSM calculations. Based on the observations in this measurement, the 0$_{5}^{+}$ state at 1774.0-keV appears to more closely align with an excitation in a spherical minimum, while the 0$_{4}^{+}$ state would be associated with a weakly-deformed shape similar to that of the 0$_{2}^{+}$ level. The experimental results support the MCSM calculations. Future studies are needed to reach more definitive conclusions about shape coexistence in $^{100}$Zr.


\makeatletter
\def\thickhline{%
  \noalign{\ifnum0=`}\fi\hrule \@height \thickarrayrulewidth \futurelet
   \reserved@a\@xthickhline}
\def\@xthickhline{\ifx\reserved@a\thickhline
               \vskip\doublerulesep
               \vskip-\thickarrayrulewidth
             \fi
      \ifnum0=`{\fi}}
\makeatother

\newlength{\thickarrayrulewidth}
\setlength{\thickarrayrulewidth}{2\arrayrulewidth}


\section{Acknowledgements}

This research used resources of Argonne National Laboratory’s ATLAS facility, which is a Department of Energy Office of Science User Facility. This material is based upon work supported by the U.S. Department of Energy, Office of Science, Office of Nuclear Physics, under Contract No. DE-AC02-98CH10886 (BNL), No. DE-AC02-06CH11357 (ANL), and grants No. DE-FG02-97ER41041 (UNC) and No. DE-FG02-97ER41033 (TUNL), the National Nuclear Security Administration, Office of Defense Nuclear Nonproliferation R\&D (NA-22) and the National Science Foundation under Grant No. PHY-1907409, and German BMBF Grant Nos. 05P21RDFN1 and 05P21RDCI2.”



\bibliographystyle{unsrt}

\end{document}